\documentclass[prd,aps,floats,twocolumn]{revtex4}
\usepackage{amssymb}
\usepackage{amsmath}
\usepackage{mathrsfs}
\usepackage{latexsym}
\usepackage[dvips]{graphicx}
\begin{document}

\title{Spinor Theory of Gravity}

\author{M. Novello}
\affiliation{
Institute of Cosmology, Relativity and Astrophysics ICRA/CBPF \\
Rua Dr. Xavier Sigaud 150, Urca 22290-180 Rio de Janeiro,
RJ-Brazil}
\date{\today}

\begin{abstract}
The proposal of this work is to provide an answer to the following
question: is it possible to treat the metric of space-time - that in
General Relativity (GR) describes the gravitational interaction - as
an effective geometry? In other words, to obtain the dynamics of the
metric tensor $g_{\mu\nu}$ as a consequence of the dynamics of other
fields. In this work we will use a slight modification of the
non-linear equation of motion of a spinor field proposed some years
ago by Heisenberg, although in a completely distinct context, to
obtain a field theory that provides a framework equivalent to the
way GR represents the gravitational interaction. In particular we
exhibit a solution of the equations of motion that represents the
gravitational field of a compact object and compare it with the
corresponding Schwarzschild solution of General Relativity.

\end{abstract}

\vskip2pc
 \maketitle

\section{Introduction}

By the Equivalence Principle \cite{will}  the gravitational
interaction may be described as a modification of the geometry of
space-time. In the General Relativity theory this idea was
implemented by assuming that there exists a unique geometrical
structure which acts on all forms of matter and energy (including
the gravitational one) in the same way. Moreover, Einstein
postulated an equation of motion to describe the evolution of the
geometry under the very natural assumption that the gravitational
field (identified with the geometrical structure) should have a
dynamics of its own. These two parts of the description are related
but independent. Taken together, the Equivalence Principle and
Einstein's equation, constitute the basis of a successful program of
a theory of gravity.

Is this the unique way to deal with the universality of
gravitational processes? In this work we propose a new way to
implement the Equivalence Principle in which the geometry acting on
matter is not an independent field, and as such does not posses its
own dynamics. Instead, it inherits one from the dynamics of two
fundamental spinor fields $\Psi$ and $\Upsilon$ which are
responsible for the gravitational interaction and from which an
effective geometry will be constructed. The nonlinear character of
gravity should be present already at the most basic level of these
fundamental structures. It seems natural to describe this
nonlinearity in terms of the invariants constructed with the spinor
fields. The simplest way to build a concrete model is to use the
standard form of a contraction of the currents of these fields, e.
g. $J_{\mu} \, J^{\mu}$ to construct the Lagrangian of the theory.
This will lead us to deal with Heisenberg's equation of motion,
which has precisely this form, although it originated in a
completely different context and is written in terms of others
invariants.

We assume that these two fields (which are half-integer
representation of the Poincar\'e group) interact universally with
all other forms of matter and energy. As a consequence, this process
can be viewed as nothing but a change of the metric of the
space-time. In other words we shall show that the influence of these
spinor fields on matter/energy is completely equivalent to a
modification of the background geometry into an effective Riemannian
geometry $g_{\mu\nu}.$ In this aspect this theory agrees with the
idea of General Relativity theory which states that the Equivalence
Principle implies a change on the geometry of space-time as a
consequence of the gravitational interaction. However, the
similarities between the Spinor Theory of Gravity and General
Relativity stop here.

To summarize let us stress the main steps of this new program.

a. There exist two fundamental spinor fields -- which we will name
$\Psi$ and $\Upsilon;$

b. These fields obey the nonlinear Heisenberg equation of motion;

c. The fields  $\Psi$ and $\Upsilon$ interact universally with all
forms of matter and energy;

d. As a consequence of this coupling with matter, this universal
interaction produces an effective metric;

e. The dynamics of the effective metric is already contained in the
dynamics of $\Psi$ and $\Upsilon:$ the metric does not have a
dynamics of its own, but inherits its evolution through its relation
with the fundamental spinors;

f.  We present a particular example of the effective metric in the
case of a compact spherically static object, like a star. We
compare it with the analogous analysis in the case of General
Relativity.

Before entering the analysis of these questions let us briefly
comment our motivation. As we shall see, the present proposal and
the theory of General Relativity have a common underlying idea:
the characterization of gravitational forces as nothing but the
effect on matter and energy of a modification of the geometry of
space-time. This major property of General Relativity, remains
unchanged. The main difference is about the dynamics that this
geometry obeys. In GR the dynamics of the gravitational field
depends on the curvature invariants; in the Spinor Theory  of
Gravity such a specific dynamics simply does not exists: the
geometry evolves in space-time according to the dynamics of the
spinors $\Psi$ and $\Upsilon.$ The metric is not a field of its
own, it does not have an independent reality but is just a
consequence of the universal coupling of matter with the
fundamental spinors. The motivation of walking down only half of
Einstein's path to General Relativity is to avoid certain known
problems that still plague this theory, including its difficult
passage to the quantum world; the questions put into evidence by
astrophysics involving many discoveries such as the acceleration
of the universe, the problems requiring dark matter, dark energy.
Up to now, only highly speculative ideas have appeared to solve
such problems, culminating with a plethora of proposals of scalar
fields with negative energies. Also, the possibility of relating
gravitational phenomena to elementary fields, usually identified
to matter fields, may present new possibilities of future
developments on a new road towards a unified program. With this in
mind, we will present a toy model to discuss such an idea.

In section II we present the mathematical background used in the
paper. In particular we analyze the standard form of the internal
connection needed to produce a covariant definition of the
derivative of a spinor. Even in a flat manifold, like Minkowski
geometry, this generalization of the derivative is needed to obtain
a covariant description in a non-Euclidean coordinate system. We
recall the idea that the expression of Fock-Ivanenko that displays
the internal connection in terms of derivatives of the metric and of
the Dirac matrices $\gamma_{\mu}$'s is not unique. The form of
Fock-Ivanenko assumes that the covariant derivative of the
$\gamma_{\mu}$'s vanishes. This possibility is allowed by the fact
that the background manifold is Riemannian, but any function of the
Clifford algebra could imply the same property for the metric
structure. We are then led to add to the internal connection a
vector $U_{\mu}.$ This vector is an element of the Clifford algebra
and will be constructed in terms of the currents of the fundamental
spinor fields. We show that this choice of the connection allows a
new geometrical interpretation of the Heisenberg self-interaction of
a spinor field. Then, in section V, we review the field theory
formulation of General Relativity as it was described in the fifties
by Gupta, Feynman \cite{Feynman} and others,  and more recently in
\cite{GPP}. In this formulation the gravitational field  can be
described alternatively either as the metric of space-time -- as in
Einstein's original version -- or as a field $\varphi_{\mu\nu}$ in
an arbitrary unobservable background geometry, which is chosen to be
minkowski (see also \cite{GPPdesitter}). We shall see that by
universally coupling the spinor fields to all forms of matter and
energy, a metric structure appears, in a similar way to the field
theoretical description of GR. The main distinction between these
two approach concerns the status of this metric. In General
Relativity it has a dynamics provided by a lagrangian constructed in
terms of the curvature invariants. In our proposal, this is not the
case. The metric is an effective way to describe gravity and it
appears because of the universal form of the coupling of
matter/energy of any form and the fundamental spinors. In section VI
we present a toy model of the dynamics of these fields. Section VII
is dedicated to a particular solution of these equations in the
description of the static and spherically symmetric external
gravitational field generated by a massive object such as a star. We
exhibit the resulting effective metric and compare it with the
similar situation in the General Relativity theory. We end with some
conclusions and future perspectives.

\section{Definitions and some mathematical machinery}
\protect\label{algebraic}

The vector and the axial currents are defined in the standard way:
\[
J^{\mu}\equiv \overline{\Psi} \gamma^{\mu}  \Psi
\]
\[
I^{\mu}\equiv \overline{\Psi} \gamma^{\mu} \gamma^{5} \Psi.
\]
In this paper we deal with two spinor fields $\Psi$ and
$\Upsilon.$ We use capital symbols to represent the currents
constructed with $\Psi$ as above and lower case to represent the
corresponding terms of the  spinor $\Upsilon,$ namely,
\[
j^{\mu}\equiv \overline{\Upsilon} \gamma^{\mu}  \Upsilon
\]
\[
i^{\mu}\equiv \overline{\Upsilon} \gamma^{\mu} \gamma^{5} \Upsilon.
\]

We use the standard convention and definitions (see, for instance
\cite{Elbaz}). For completeness we recall:
\[
\bar{\Psi} \equiv  \Psi^{+} \gamma^{0}.
\]
The $\gamma^{5}$ is hermitian and the other $\gamma_{\mu}$ obey the
hermiticity relation
\[
\gamma_{\mu}^{+} = \gamma^{0} \gamma_{\mu} \gamma^{0}.
\]

The properties needed to analyse non-linear spinors are contained
in the Pauli-Kofink (PK) relation. These are identities that
establish a set of tensor relations concerning elements of the
four-dimensional Clifford algebra. The main property states that,
for any element $Q$ of this algebra, the PK relation ensures the
validity of the identity:
\begin{equation}
(\bar{\Psi} Q \gamma_{\lambda} \Psi) \gamma^{\lambda} \Psi  =
(\bar{\Psi} Q \Psi)  \Psi  -  (\bar{\Psi} Q \gamma_{5} \Psi)
\gamma_{5} \Psi. \protect\label{H5}
\end{equation}
for $Q$ equal to $\mathbb{I}$ , $\gamma^{\mu}$, $\gamma_{5}$ and
$\gamma^{\mu} \gamma_{5},$ where $\mathbb{I}$ is the identity of the
Clifford algebra. As a consequence of this relation we obtain two
extremely important facts:
\begin{itemize}
 \item{The norm of the currents $J_{\mu}$ and $I_{\mu}$ have the same
value and opposite sign.}
 \item{The vectors  $J_{\mu}$ and $I_{\mu}$ are orthogonal.}
\end{itemize}

Indeed, using the Pauli-Kofink relation we have, for $Q =
\mathbb{I}$
\[
(\bar{\Psi}  \gamma_{\lambda} \Psi) \gamma^{\lambda} \Psi  =
(\bar{\Psi}  \Psi)  \Psi  -  (\bar{\Psi} \gamma_{5} \Psi) \gamma_{5}
 \Psi.
\]
Multiplying by $\bar{\Psi}$ and using the above definitions yields
\begin{equation}
J^{\mu} J_{\mu} = A^{2} +  B^{2}, \protect\label{H6}
\end{equation}
where $A \equiv  \bar{\Psi} \, \Psi$ and   $B \equiv i \bar{\Psi}
\, \gamma^{5} \Psi.$
 We also have
\begin{displaymath}
(\bar{\Psi} \gamma_{5} \gamma_{\lambda} \Psi) \gamma^{\lambda} \Psi
= (\bar{\Psi} \gamma_{5} \Psi)  \Psi  - (\bar{\Psi} \Psi) \gamma_{5}
\Psi.
\end{displaymath}
From which it follows that the norm of $I_{\mu}$ is
\begin{equation}
I^{\mu} I_{\mu} = - A^{2} - B^{2} \protect\label{H7}
\end{equation}
and that the four-vector currents are orthogonal
\begin{equation}
I_{\mu} J^{\mu} = 0. \protect\label{H71}
\end{equation}
It follows that the current $J_{\mu}$ is a time-like vector; and
the axial current is space-like.

\subsubsection{Internal connection}

It is useful to treat the equation of motion of the fundamental
spinors in a non-euclidean system of coordinates. In order to deal
with the covariance of the theory it is necessary to introduce the
concept of internal connection. In the case of an arbitrary
Riemannian geometry (of which the Minkowski metric is a particular
case) Fock and Ivanenko displayed the main properties needed to
obtain such covariant description in the case of a spinor. This
means, exchanging the simple derivative for a covariant one
defined by
\begin{equation}
\nabla_{\mu} \Psi = \partial_{\mu} \Psi - i \Gamma_{\mu} \Psi.
\label{11agosto1720}
\end{equation}
In the same way the elements of the Clifford algebra must be
conveniently modified. Suppose that we are dealing with the
Minkowski geometry in a spherical coordinate system, as is the
case later on. The metric takes the form
\begin{equation}
ds^{2} = dt^{2} - dr^{2} - r^{2} d\theta^{2} - r^{2} sin^{2}\theta
d\varphi^{2}. \label{11agosto1725}
\end{equation}
In consequence, the $\gamma_{\mu}$'s are given in terms of the
constant $\widetilde{\gamma}_{\mu}$ as follows:
\begin{eqnarray}
\gamma_{0} &=&  \widetilde{\gamma}_{0} \nonumber \\
\gamma_{1} &=& \widetilde{\gamma}_{1}  \nonumber \\
\gamma_{2} &=&  r \, \widetilde{\gamma}_{2} \nonumber \\
\gamma_{3} &=&  r \, sin\theta \, \widetilde{\gamma}_{3}.
\nonumber
\end{eqnarray}
For later use we display our convention of the constant
$\gamma_{\mu}$'s:
\[
\widetilde{\gamma}^{0} =  \left(
\begin{array}{cccc}
I_{2} &
   0 \nonumber \\
 0 &
  -\, I_{2} &
    \nonumber \\
  \end{array}
\right) \nonumber
\]

\[
\widetilde{\gamma}_{k} =  \left(
\begin{array}{cccc}
0 &
   \sigma_{k} \\
 -\sigma_{k} &
  0 &
    \\
  \end{array}
\right) \nonumber
\]
\[
\gamma^{5} =  \left(
\begin{array}{cccc}
0 &
   I_{2}  \\
 I_{2}&
  0 &
    \\
  \end{array}
\right). \nonumber
\]

This form was obtained by using the property
\begin{equation}
\gamma_{\mu} \, \gamma_{\nu} + \gamma_{\nu} \, \gamma_{\mu} = 2
g_{\mu\nu} \, \mathbb{I}.
\label{11agosto20}
 \end{equation}  Note
that from now on we will write simply $1$ to represent the
identity of the Clifford algebra.

In the case of the original Fock-Ivanenko condition (i.e.,
vanishing of the covariant derivative of the $\gamma_{\mu}$) one
obtains the form for the FI connection:
\begin{equation}
\Gamma_{\mu}^{0} = \frac{1}{8} \, \left[ \gamma^{\alpha}
\gamma_{\mu \, ,\alpha} - \gamma_{\mu \, ,\alpha} \gamma^{\alpha}
+ \Gamma^{\epsilon}_{\mu\nu} \, (\gamma_{\epsilon} \gamma^{\nu} -
\gamma^{\nu} \gamma_{\epsilon}) \right]. \label{11agosto2940}
\end{equation}
The index $0$ in $\Gamma_{\mu}$ is just a reminder that we are
dealing with a Minkowski background in an arbitrary system of
coordinates. We can globally annihilate  such connection by moving
to an Euclidean constant coordinate system.

\subsubsection{Generalized internal connection}
The expression of the internal connection as displayed by Fock and
Ivanenko was obtained by assuming that the covariant derivative of
all $\gamma_{\mu}$ vanish. This is a direct consequence of
relation (\ref{11agosto20}). Indeed, $\nabla_{\mu} \, \gamma_{\nu}
= 0$ implies that the metric is Riemannian: $ \nabla_{\mu} \,
g_{\alpha\beta} = 0.$ However, although the condition of vanishing
covariant derivatives of $\gamma_{\mu}$ is enough to guarantee the
Riemannian structure of the geometry, it is not necessary. In
\cite{Novello70} a case is examined in which the dynamics of the
Clifford structure is driven by the condition of the commutator:
\begin{equation}
\nabla_{\mu} \, \gamma_{\nu} =  [ U_{\mu} \, , \gamma_{\nu}],
\label{11agosto2025}
\end{equation}
where $U_{\mu}$ is an arbitrary element of the Clifford algebra.

Indeed, from the relation (\ref{11agosto20}) and using the above
expression with $U_{\mu} = A_{\mu} + B_{\mu} \gamma_{5},$ we have
for arbitrary vectors $A_{\mu}$ and $B_{\mu}:$
\begin{equation}
\nabla_{\mu} \, \gamma_{\nu} =  [ A_{\mu} + B_{\mu} \,\gamma_{5}
 , \gamma_{\nu}]. \label{26agosto9025}
\end{equation}
We have
\[
\nabla_{\mu} \, g_{\alpha\beta} = [ U_{\mu} , \gamma_{\alpha} ] \,
\gamma_{\beta} +  \gamma_{\alpha} [ U_{\mu} , \gamma_{\beta} ] \,
+ [ U_{\mu} , \gamma_{\beta} ] \, \gamma_{\alpha} +
 \gamma_{\beta} \, [ U_{\mu} , \gamma_{\alpha} ]
\]
and using the property that $\gamma_{5}$ anti-commutes with all
$\gamma_{\nu},$ it follows that $ \nabla_{\mu} \, g_{\alpha\beta}=
0.$ This holds for arbitrary vectors $A_{\mu}$ and $B_{\mu}.$

We shall see that the internal connection obtained in this way
provides an equivalent way to describe the non linear structure of
Heisenberg spinors for a convenient choice of $U_{\mu}.$  Thus the
internal connection takes the form
\begin{equation}
\Gamma_{\mu} =  \Gamma_{\mu}^{0} - i U_{\mu}. \label{18agosto910}
\end{equation}

\section{HEISENBERG QUARTIC SELF-INTERACTING SPINORS}
\protect\label{h-FERMIONS}

\subsubsection{Historical comment}

 In a series of papers ( see \cite{heisenberg1} for a complete list)
 Heisenberg examined a
proposal regarding a complete quantum theory of fields and
elementary particles. Such a huge and ambitious program did not
fulfill his initial expectation. It is not our intention here to
discuss this program. For our purpose, it is important only to
retain the original non linear equation of motion which Heisenberg
postulated for the constituents of the fundamental material blocks
of all existing matter. The modern point of view has developed in a
very different direction and it is sufficient to take a look at the
book of Particle Data Properties \cite{Caso} and the description of
our actual knowledge of the elementary particle properties to
realize how far from Heisenberg dream the theory has gone.

So much for the historical context. What we would like to retain
from Heisenberg's approach reduces exclusively to his suggestion
of a non linear equation of motion for a spinor field. We will use
this equation for both our fundamental spinors, once as we will
now see, it is the simplest non-linear dynamics that can be
constructed in a covariant way. Let $\Psi$ and $\Upsilon$ be the
fundamental four-component spinor field. The dynamics of $\Psi$
(resp., $\Upsilon$) is given by the self-interaction Lagrangian
(we are using the conventional units were $\hbar = c = 1):$
\begin{equation}
L = \frac{i}{2} \bar{\Psi} \gamma^{\mu} \partial_{\mu} \Psi -
\frac{i}{2} \partial_{\mu} \bar{\Psi} \gamma^{\mu} \Psi - V(\Psi).
\protect\label{H1}
\end{equation}
The potential $V$ is constructed with the two scalars that can be
formed with $\Psi$, that is $A$ and $B.$ We will only consider the
 Heisenberg potential that is
\begin{equation}
V = s \left( A^{2} + B^{2} \right)
 \protect\label{H3}
\end{equation}
where $s$ is a real parameter of dimension $(length)^{2}.$

 This potential can be written in an equivalent and more suggestive form in terms
of the associated currents $J_{\mu}$ and $I_{\mu}.$ These two
vectors will become the basic ingredients of the model which we
will deal with in the present paper. As we have anticipated above
in equation (\ref{H6}), the Heisenberg potential $V$ is nothing
but the norm of the four-vector current $J^{\mu}$.

The Heisenberg non linear equation of motion:
\begin{equation}
i \gamma^{\mu} \partial_{\mu} \Psi - 2s (A + i B \gamma_{5}) \Psi
= 0, \protect\label{H8}
\end{equation}
follows from the Lagrangian (\ref{H1}) with the potential $V = s
J^{2}.$

\section{Geometrical realization of the Heisenberg spinor}
In this section we show how to understand the self-coupling of
equation (\ref{H8}) in terms of a modification of the internal
connection. In so doing, we are preparing our analysis for the
universal gravitational interaction of the non-linear spinor
theory. Let us use the form (\ref{18agosto910}) and set
\begin{equation}
\Gamma_{\mu} = -i \left( a J_{\mu} + b I_{\mu} \right) \left(
\mathbb{I} + \gamma^{5} \right) \label{11agosto2050}
\end{equation}
in which, for simplicity we use an Euclidean coordinate system in
which the Fock-Ivanenko standard part of the connection vanishes.
Thus, the Lagrangian of the fundamental spinor takes the form
\begin{eqnarray}
L &=&  \frac{i}{2} \bar{\Psi} \gamma^{\mu} \nabla_{\mu} \Psi -
\frac{i}{2} \, \nabla_{\mu} \bar{\Psi} \gamma^{\mu} \Psi \nonumber
\\
&=&  \frac{i}{2} \bar{\Psi} \gamma^{\mu} \partial_{\mu} \Psi +
\frac{1}{2} \, \bar{\Psi} \gamma^{\mu} \Gamma_{\mu} \Psi + h.c.
\protect\label{11agosto2051}
\end{eqnarray}
Substituting the form (\ref{11agosto2050}) in this Lagrangian  we
obtain
 \begin{equation}
L = \frac{i}{2} \bar{\Psi} \gamma^{\mu} \partial_{\mu} \Psi -
\frac{i}{2} \partial_{\mu} \bar{\Psi} \gamma^{\mu} \Psi -
\frac{i}{2} \, \left[ (a - \bar{a}) - ( b - \bar{b} \right] \,
J_{\mu} \, J^{\mu}. \label{11agosto21}
\end{equation}
This is precisely the expression of Heisenberg Lagrangian
(\ref{H1}) which led us to the identification
\begin{equation}
s = \frac{i}{2} \, \left[ (a - \bar{a}) - (b - \bar{b}) \right].
\label{13agosto15}
\end{equation}
Thus we succeeded to present Heisenberg self-interaction as a
modification  of the internal connection structure.

\subsubsection{Two fundamental spinors}
Our theory contains two spinors that obey Heisenberg type of
equations of motion. Once we can describe such self-coupling in
terms of a modification of the internal connection, this procedure
automatically implies a direct interaction between $\Psi$ and
$\Upsilon.$ We have
\begin{equation}
U_{\mu} = \left[ a (J_{\mu} + j_{\mu}) + b (I_{\mu} + i_{\mu})
\right] \left( 1 + \gamma^{5} \right). \label{17agosto2050}
\end{equation}

Besides the self-interaction terms there appears a direct
interaction between the two spinors given by

\begin{eqnarray}
L_{int} &=& \frac{i}{2} \, \left[ (a - \bar{a}) ( J_{\mu} +
j_{\mu}
) + (b - \bar{b}) ( I_{\mu} + i_{\mu} ) \right] \nonumber \\
 &&  \left[ J^{\mu} + j^{\mu} +  I^{\mu} + i^{\mu} \right].
 \nonumber
\end{eqnarray}
We set
\begin{equation}
b - \bar{b} = \beta \, (a - \bar{a}). \label{18agosto700}
\end{equation}
Thus, for the total interaction Lagrangian we find
\begin{eqnarray}
L_{int} &=& \frac{i}{2} \,  (a - \bar{a}) (1 - \beta)( J_{\mu}
J^{\mu} + j_{\mu} j^{\mu} ) \nonumber \\
&+& i \, (a - \bar{a}) \{  J_{\mu} j^{\mu} + \beta I^{\mu}i_{\mu}
\}
\nonumber \\
&+& \frac{i}{2} \,  (a - \bar{a}) (1 + \beta) (J^{\mu}i_{\mu} +
I^{\mu} j_{\mu} ). \label{18agosto701}
\end{eqnarray}

The first two terms represent the Heisenberg self interactions and
the other terms the interaction between  $\Psi$ and $\Upsilon.$ It
seems worthwhile to remark that in case $\beta = 1,$ the
Heisenberg terms vanishes and the interaction assumes the reduced
form
\begin{equation*}
L_{F} =  g_{F} \, \overline{\Psi} \gamma^{\mu} ( 1 + \gamma^{5} )
\Psi
 \, \, \, \overline{\Upsilon} \gamma_{\mu} (1 + \gamma^{5} )
 \Upsilon. 
\end{equation*}
where $ g_{F} \equiv  (\hbar c ) i (a - \bar{a}).$

\section{The universal coupling: gravity}

Half a century has already elapsed since the idea of dealing with
the content of General Relativity in terms of a field theory
propagating in a non-observable Minkowski background was presented
by Gupta, Feynman and others. In recent times this approach has
been revised and commented (see \cite{GPP} and references
therein).

The field theoretical approach goes back to the fact that Einstein
dynamics of the curvature of the Riemannian metric of space-time
can be obtained as a sort of iterative process, starting from a
linear theory of a symmetric second order tensor
$\varphi_{\mu\nu}$ and by an infinite sequence of self-interacting
process leading to a geometrical description. The definition of
the metric is provided in terms of the metric of the background
$\eta_{\mu\nu}$ as follows
\begin{equation}
g_{\mu\nu} \equiv \eta_{\mu\nu} +  \varphi_{\mu\nu}
\label{10agosto1}
\end{equation}
Note that this is not an approximation formula but an exact one.
The inverse metric $(g_{\mu\nu})^{-1} \equiv g^{\mu\nu}$ is
defined by $g_{\nu\mu}g^{\mu\alpha} = \delta^{\alpha}_{\nu}.$
Other definitions were also used, for instance,
\begin{equation*}
\sqrt{-g} \,g^{\mu\nu} \equiv \sqrt{-\gamma} \left(
\gamma^{\mu\nu} + \varphi^{\mu\nu} \right) \nonumber
\end{equation*}
where $ \gamma_{\mu\nu}$ is the background Minkowski metric, written
in an arbitrary system of coordinates (see for instance \cite{GPP}
for an analysis of the convenience of these alternative
non-equivalent definitions).

Although these theories can be named " field theories" they
contain the same metric content of General Relativity,  disguised
in a non geometrical form. The framework which will be discussed
here is totally different. It is important to emphasize that we
are not presenting a dynamics for the metric in the sense of such
field theories. Instead, the geometry is understood as an
effective one, in the sense that it is the way gravity appears for
all forms of matter and energy. However, its evolution is given by
the fundamental spinor fields $\Psi$ and $\Upsilon$. We learn from
these field theories of gravitation the way to couple the tensor
field $\varphi_{\mu\nu}$ with matter terms in order to guarantee
that the net effect of this interaction produces the modification
of the metric structure. This idea will guide us when coupling the
two fundamental spinors with all forms of matter and energy in
order to arrive at the same equivalent interpretation of the
identification of the gravitational field with the metric of the
space-time.

\section{The universal coupling of $\Psi$ and $\Upsilon$ with
matter}

From the previous section, the reader understands that our strategy
is to treat the interaction of the spinors fields in terms of a
modification of an internal connection. Now we face the question:
how does matter of any form and any kind of energy interact with
these two fields? Following this strategy we make a major hypothesis
(which substitutes the corresponding hypothesis made by Einstein on
the dynamics of $g_{\mu\nu}$) that the spinors interact universally
with all forms of matter/energy through the modification of the
internal connection $\Gamma_{\mu}.$ Let us review briefly the way GR
describes this coupling and compare it with our procedure.

Let $L_{0}$ be the Lagrangian of a certain matter distribution in
the absence of gravitational forces given by
\begin{equation*}
S_{0} = \int \sqrt{-\gamma} \, L_{0} = \int \sqrt{-\gamma} \,
B^{\mu\nu} \, \gamma_{\mu\nu}.
\end{equation*}

 In order to couple the matter, described by this Lagrangian, with gravity
the procedure in the field theory formulation of General
Relativity is made through the use of the Equivalence Principle
or, as sometimes it is named, the minimal coupling principle. This
means substituting of all the terms in the action $S_{0}$ in which
the Minkowski metric $\gamma_{\mu\nu}$ appears by the general
metric $g_{\mu\nu}$ and its inverse $g^{\mu\nu}.$  Let us give two
examples. First we consider the case of a scalar field $\Phi.$ In
the Minkowski background its dynamics is provided by
\begin{equation*}
S_{0} = \int \sqrt{-\gamma} \, \partial^{\mu}\Phi \,
\partial^{\nu}\Phi \, \gamma_{\mu\nu}.
\end{equation*}

In this case $B^{\mu\nu}$ can be
written in terms of the energy-momentum tensor defined as
\begin{equation*}
T_{\mu\nu} = \frac{2}{\sqrt{-\gamma}} \, \frac{\delta}{\delta
\gamma^{\mu\nu}} \left( \sqrt{-\gamma} L \right).
\end{equation*}
Indeed, a direct calculation yields
\begin{equation*}
T^{\mu\nu} = \partial_{\alpha}\Phi \partial_{\beta}\Phi \,
\gamma^{\alpha\mu} \gamma^{\beta\nu} - \frac{1}{2}
\partial_{\lambda}\Phi \partial_{\sigma}\Phi \,
\gamma^{\lambda\sigma} \gamma^{\mu\nu}
\end{equation*}
immediately implying the expression
\begin{equation*}
B^{\mu\nu} = T^{\mu\nu} - \frac{1}{2} T \gamma^{\mu\nu},
\end{equation*}
where $ T \equiv T^{\mu\nu} \, \gamma_{\mu\nu}.$
 The corresponding action, including the gravitational
 interaction, is obtained by changing all $\gamma_{\mu\nu}$ and its inverse
$\gamma^{\mu\nu}$ with the corresponding $g_{\mu\nu} =
\gamma_{\mu\nu} + \varphi_{\mu\nu}$ which yields
\begin{equation*}
S = \int \sqrt{-g} \, \partial^{\mu}\Phi \,
\partial^{\nu}\Phi \, g_{\mu\nu},
\end{equation*}
where $g = det \, g_{\mu\nu}.$ In this case
\begin{equation*}
B^{\mu\nu} = \frac{\sqrt{-g}}{\sqrt{-\gamma}} \, \, [T^{\mu\nu} -
\frac{1}{2} T g^{\mu\nu}].
\end{equation*}

 Let us consider now the case of
the electromagnetic field. The action is given by
\begin{equation*}
S_{0} = \int \sqrt{-\gamma} \, F^{\alpha\mu} \, F^{\beta\nu} \,
\gamma_{\alpha\beta} \, \gamma_{\mu\nu}
\end{equation*}
The same reasoning yields
\begin{equation*}
B^{\mu\nu} \equiv  F^{\alpha\mu} \, F^{\beta\nu} \,
\gamma_{\alpha\beta}
\end{equation*}
and we follow the same rules as in the previous scalar equation of
motion. The fact that it is not possible to write the tensor
$B^{\mu\nu}$ in terms of the energy-momentum tensor in this case is
due to the fact that this tensor is traceless.

Let us turn now to the analogous analysis in the Spinor Theory of
Gravity. We follow a similar procedure. Our strategy is to modify
the internal connection and change $U_{\mu}$ by the form:
\begin{equation}
U_{\mu} = U_{\mu}^{1} + U_{\mu}^{2} \label{1setembro 1400}
\end{equation}
where

\[
U_{\mu}^{1} =  \left[ a (J_{\mu} + j_{\mu}) + b (I_{\mu} + i_{\mu})
\right] \,  ( 1 + \gamma_{5})
\]
and \[
 U_{\mu}^{2} = - \frac{\lambda}{\sqrt{X}} \, \left[ a
(J_{\alpha} + j_{\alpha}) + b (I_{\alpha} + i_{\alpha}) \right] \,
B_{\mu}^{\alpha} ( 1 + \gamma_{5} )
\]
where $X$ is the sum of the norms of the vectors $J_{\mu}$ and
$j_{\mu},$ i.e. $X = J_{\mu}J^{\mu} + j_{\mu}j^{\mu}$ and
$\lambda$ is a constant of dimension $(energy)^{-1}.$ The first
term $ U_{\mu}^{1}$ represents the free-field and the second one
the interaction of the fundamental spinors with matter.

In the present theory this is how matter affects the spinor fields.
To be explicit, let us write the complete Lagrangian for one of the
spinors (analogous form corresponds to the other one). Taking into
account the fact that $B_{\mu\nu}$ is symmetric and inserting
$U_{\mu}^{2}$ into equation (\ref{11agosto2051}), we find for the
matter interaction part
\begin{equation}
L_{g} = - g_{F} \, \frac{\lambda}{4} \,  \frac{\left( c_{\mu\nu} +
c_{\nu\mu} \right)}{\sqrt{X}} \, B^{\mu\nu} \label{18agosto925}
\end{equation}
where
\begin{equation*}
c_{\mu\nu} = [ J_{\mu} + j_{\mu} +  I_{\mu} + i_{\mu} ] [ J_{\nu} +
j_{\nu} +  \beta \, (I_{\nu} + i_{\nu}) ]
\end{equation*}

This form of interaction of matter/energy with the fundamental
spinor fields leads to the definition of a metric, in the same way
as in the field theory representation of General Relativity namely
\begin{equation}
g_{\mu\nu} = \eta_{\mu\nu} + \varphi_{\mu\nu} \label{18agosto940}
\end{equation}
where the field $\varphi_{\mu\nu}$ is chosen to be non-dimensional,
that is, we set:
\begin{equation}
\varphi_{\mu\nu} = - \frac{g_{F} \, \lambda}{4} \,
\frac{1}{\sqrt{X}} \, (c_{\mu\nu} + c_{\nu\mu}) \label{18agosto945}
\end{equation}


Due to our choice of the definition of the effective metric
$g_{\mu\nu}$  it follows that $B^{\mu\nu}$ is the same as in the
field theory representation of General Relativity presented above.
This way of coupling matter with the fundamental spinors
guarantees that in what concerns the behavior of matter the
analysis of General Relativity is still valid in the present
theory: free particles follow geodesics in the effective geometry
$g_{\mu\nu.}$  The most important task now is to analyze the
consequences of this theory. We start this by studying the
effective metric generated by the gravitational process in the
neighborhood of a massive object, like a star.

\section{Gravitational field of a compact object}

In the absence of matter and energy, the effective metric can be
obtained by a direct solution of the Heisenberg equation and the
identification of $g_{\mu\nu}$ through equation
(\ref{18agosto940}). The equations of motion in this case are
\begin{eqnarray}
i \gamma^{\mu} \, \partial_{\mu} \Psi &+&  \gamma^{\mu} \,
\Gamma_{\mu}^{(0)} \Psi \nonumber \\
&-& 2s ( A + i B \gamma^{5} ) \Psi \nonumber \\
&-&  g_{F}  \,
 \gamma^{\mu} \, \left(j_{\mu} + \frac{1 + \beta}{2} i_{\mu}\right)
 \Psi \nonumber \\
&-&  g_{F}  \, \left( \beta i_{\mu} + \frac{ 1 + \beta}{2} \,
j_{\mu} \right) \gamma^{\mu} \gamma^{5} \,  \Psi  = 0.
 \label{13agosto1510}
\end{eqnarray}
\begin{eqnarray}
i \gamma^{\mu} \, \partial_{\mu} \Upsilon &+&  \gamma^{\mu} \,
\Gamma_{\mu}^{(0)} \Upsilon \nonumber \\
&-& 2s ( \widehat{A} + i \widehat{B} \gamma^{5} ) \Upsilon \nonumber \\
&-& g_{F} \,
 \gamma^{\mu} \, \left(J_{\mu} + \frac{1 + \beta}{2} I_{\mu}\right)
 \Upsilon \nonumber \\
&-& g_{F}  \, \left( \beta I_{\mu} + \frac{1 + \beta}{2} \, J_{\mu}
\right) \gamma^{\mu} \gamma^{5} \,  \Upsilon   = 0.
 \label{13agosto1510}
\end{eqnarray}
where $\widehat{A} \equiv \bar{\Upsilon} \, \Upsilon$ and
$\widehat{B} \equiv i \bar{\Upsilon} \, \gamma^{5} \Upsilon.$

 This is a highly non linear system that must be solved in order
 to obtain the effective metric. We succeeded in finding a
  solution in the case of a spherically symmetric and static
  configuration. Using the background Minkowski metric in the form
(\ref{11agosto1725}) we obtain the unique non identically
background FI connection:
\begin{eqnarray*}
\Gamma_{2}^{(0)} &=& \frac{1}{2} \, \widetilde{\gamma}_{1} \,
\widetilde{\gamma}_{2} \nonumber \\
\Gamma_{3}^{(0)} &=& \frac{1}{2} \, sin\theta \,
\widetilde{\gamma}_{1} \, \widetilde{\gamma}_{3} \, + \frac{1}{2}
\, cos\theta \, \widetilde{\gamma}_{2} \, \widetilde{\gamma}_{3}
\nonumber
\end{eqnarray*}

We will look for a solution of the form
\begin{equation*}
\Psi =  f(r) \, e^{i\varepsilon ln r} \, e^{i h(\theta)} \, \Psi^{0}
\end{equation*}
\begin{equation*}
\Upsilon =   g(r) \, e^{i\tau ln r} \, e^{i l(\theta} \,
\Upsilon^{0} 
\end{equation*}
where $\varepsilon$ and $\tau$ are constants; $\Psi^{0}$ and
$\Upsilon^{0}$ are constant spinors. The Heisenberg equation of
motion is solved if $h(\theta)$ and $l(\theta)$ are proportional to
$ln \sqrt{sin \theta}.$ Moreover, $f(r)$ and $g(r)$ obey the
equations
\begin{equation*}
\frac{1}{f^{3}} \, \frac{d f}{dr} = constant,
\end{equation*}
\begin{equation*}
\frac{1}{g^{3}} \frac{dg}{dr}= constant.
\end{equation*}
We then have
\begin{equation}
\Psi = \frac{u}{\sqrt{r}} \, e^{i\varepsilon ln r}\, e^{i ln
\sqrt{sin\theta}} \, \Psi^{0}, \label{14agosto1415}
\end{equation}
\begin{equation}
\Upsilon = \frac{u'}{\sqrt{r}} \, e^{i \tau ln r} \, e^{i ln
\sqrt{sin\theta}} \, \Upsilon^{0}. \label{14agosto1415}
\end{equation}
for constants $u$ and $u'.$  The dependence on the angle $\theta$
disappears in both (vector and axial) currents. The
$r^{-\frac{1}{2}}$ term depends on the fact that the Heisenberg
potential is of quartic order. Any other dependence should yield a
different functional dependence for the effective metric. As we
shall see next, this form is crucial in order to obtain the good
behavior of the metric in the newtonian limit.

We set
\begin{equation}
\Psi^{0} =  \left(
\begin{array}{c}
\varphi^{0} \\
 \eta^{0}.
  \end{array}
\right)
\end{equation}

 To solve the equation of motion, the constant
spinor $\Psi^{0}$ (correspondingly $\Upsilon^{0})$ must satisfy a
set of equations. We set
\begin{equation} \varphi^{0} = ( c_{1} +
c_{2} \, \sigma_{1} ) \, \eta^{0} \label{21agosto1420}
\end{equation}
We look for a solution such that
$$ \sigma_{1}  \, \eta^{0} =  \eta^{0}. $$
which yields
$$ \varphi^{0} = i \, R  \, \eta^{0}, $$
where $R$ is a real number. Note that all currents from the
expression of $ \Psi$ and $\Upsilon$ are of the form $a^{\mu}/r$ for
different constant vectors $a^{\mu}. $  After a rather long and
tedious calculation we obtain the final expressions of these
currents constructed with our solution. It is precisely these
currents that provide the effective metric, namely:
\begin{eqnarray*}
J_{o} &=& \frac{p}{r} \nonumber \\
I_{o} &=& \frac{q}{r} \nonumber \\
J_{1} &=& \frac{m}{r} \nonumber \\
I_{1} &=& \frac{n}{r} \nonumber \\
\end{eqnarray*}
and analogous formulas for the spinor $\Upsilon:$
\begin{eqnarray*}
j_{o} &=& \frac{p'}{r} \nonumber \\
i_{o} &=& \frac{q'}{r} \nonumber \\
j_{1} &=& \frac{m'}{r} \nonumber \\
i_{1} &=& \frac{n'}{r} \nonumber \\
\end{eqnarray*}
where
\begin{eqnarray*}
p &=& [c_{1}\bar{c}_{1} + c_{2}\bar{c}_{2} + 1 ]  \eta^{+} \eta  +
 [c_{1}\bar{c}_{2} + c_{2}\bar{c}_{1} ]   \eta^{+} \sigma_{1} \eta   \nonumber \\
q &=& - [c_{1} + \bar{c}_{1} ] \eta^{+} \eta -
 [c_{2} + \bar{c}_{2} ]   \eta^{+} \sigma_{1} \eta   \nonumber \\
m &=& [c_{2} + \bar{c}_{2}]  \eta^{+} \eta   +
 [c_{1} + \bar{c}_{1} ]   \eta^{+} \sigma_{1} \eta  \nonumber \\
n &=&  [c_{1}\bar{c}_{2} + c_{2}\bar{c}_{1} ]   \eta^{+} \eta  +
[c_{1}\bar{c}_{1} + c_{2}\bar{c}_{2} + 1 ]  \eta^{+} \sigma_{1}
\eta \nonumber
\end{eqnarray*}
Similar formulas holds for the corresponding quantities
constructed with $\Upsilon$ involving $p', q', m', n'.$
Analogously we set
\begin{equation}
\Upsilon^{0} =  \left(
\begin{array}{c}
\chi^{0} \\
 \zeta^{0}
  \end{array}
\right)
\end{equation}
and:
\begin{equation}
\chi^{0} = ( d_{1} + d_{2} \, \sigma_{1} ) \, \zeta^{0}
\label{21agosto1420}
\end{equation}
where $$\sigma_{1}  \, \zeta^{0} = - \, \zeta^{0},$$ and

$$ \chi^{0} = i \, S \, \zeta^{0}$$
where $S$ is a real number.

Since the constants $c_{1}, c_{2}, d_{1}$ and $ d_{2}$ are purely
imaginary numbers it follows that $m = q = m'= q'= 0.$ Consistency
imposes the four conditions
\begin{equation}
\varepsilon =  1 - \frac{g_{F}}{2} \, (1 + \beta ) ( n' + p'),
\label{29agosto725}
\end{equation}
\begin{equation}
\tau =  1 - \frac{g_{F}}{2} \, (1 + \beta ) (n + p)
 \label{29agosto730}
\end{equation}
\begin{equation}
 - 2s   \, u^{2} A_{0}  +  2s R \, u^{2} \,  B_{0} + g_{F} (p' + \beta n') - \frac{1}{2} R = 0
  \label{24agosto10}
\end{equation}
\begin{equation}
 2s \, (u)^{2} ( R {A}_{0} + B_{0} ) +  R  g_{F} \, ( p' + \beta n') + \frac{1}{2} = 0.
\label{23agosto755}
\end{equation}

  By symmetry, the components $(2)$ and $(3)$ of the currents $J_{\mu},
I_{\mu}, j_{\mu}, i_{\mu}, $ must vanish. This is possible if the
constant spinors satisfy:
\begin{eqnarray}
\eta^{+}_{0} \sigma_{2} \eta_{0} &=& 0 \nonumber \\
\eta^{+}_{0} \sigma_{3} \eta_{0} &=& 0 \label{21agosto1520}
\end{eqnarray}
and
\begin{eqnarray}
\zeta^{+}_{0} \sigma_{2} \zeta_{0} &=& 0 \nonumber \\
\zeta^{+}_{0} \sigma_{3} \zeta_{0} &=& 0 \label{24agosto915}
\end{eqnarray}

Once all these conditions are satisfied there remains two arbitrary
conditions to be fixed, for instance  $\eta^{+}_{0} \, \eta_{0}$ and
$\zeta^{+}_{0} \zeta_{0}.$   Different choices yield different
solutions for the spinor fields and consequently distinct
configurations for the observable metric.

\section{The effective metric}
\label{Effective}

From the above solution of the spinor fields we can evaluate the
currents and the effective geometry that acts on all forms of
matter and energy. From its dependence on $r$ and $\theta$ we have
 that all currents depend only on $1/r.$ Using the expression of
the effective metric in terms of the spinorial fields provided by
equations (\ref{18agosto940}) and (\ref{18agosto945}), a direct
calculation gives:
\begin{eqnarray}
ds^{2} &=& ( 1 - \frac{r_{H}}{r} )  dt^{2} + 2 \frac{N}{r} dr dt
\nonumber \\
&-& ( 1 + \frac{Q}{r}) dr^{2} - r^{2} d\theta^{2} - r^{2}
sin^{2}\theta d\varphi^{2}, \label{13agosto1600}
\end{eqnarray}
where:
\begin{eqnarray}
r_{H} = \frac{g_{F} \lambda}{2} \, \frac{1}{\sqrt{Z}} (p + p')^{2}, \nonumber \\
Q = \frac{ g_{F} \lambda}{2} \, \frac{1}{\sqrt{Z}} \,  \beta \, ( n + n' )^{2}, \nonumber \\
N = - \, \frac{g_{F} \lambda}{4} \, \frac{1}{\,\sqrt{Z}} \, (p +
p')( n + n'). \nonumber
\end{eqnarray}
The constant $Z$ is defined in terms of the norm of the currents
as $ Z = X \, r^{2} = p^{2} + (p') ^{2}.$

 In order to compare this geometry with the corresponding
solution in General Relativity, we make a coordinate
transformation to eliminate the crossing term  $dr dt.$ Setting
\begin{equation*}
dt = dT +  \frac{N}{ r - r_{H}} \, dr,
\nonumber
\end{equation*}
we obtain
\begin{eqnarray}
ds^{2} &=& ( 1 - \frac{r_{H}}{r} )  dT^{2} \nonumber \\
&-& \left( 1 - \frac{r_{H}}{r} \right)^{-1} \, \left( 1 -
\frac{r_{H} - Q}{r} -
\frac{Q r_{H} - N^{2}}{r^{2}} \right)  dr^{2} \nonumber \\
 &-& r^{2} d\theta^{2} - r^{2} sin^{2}\theta d\varphi^{2}. \label{12agosto1610}
\end{eqnarray}

At this point we remark that in the case of General Relativity,
Birkhoff's theorem forbids the existence of more than one arbitrary
constant in the Schwzarschild solution. In the present case of the
Spinor Gravity theory, this theorem does not apply. Thus we can
understand the fact that this solution contains one additional
arbitrary constant. Observations \cite{will2} impose that for small
values of $r_{H}/r$ the factors $g_{00}$ and $g_{11}$ must be in the
first order respectively $g_{00} = 1 -r_{H}/r $ and $g_{11} = - 1 -
r_{H}/r.$ This fact imply that the the constants $\eta^{+}_{0} \,
\eta_{0}$ and $\zeta^{+}_{0} \, \zeta_{0}$ must be chosen such that
$r_{H} = Q. $  This fixes one constant. The other constant is
provided, as in the similar procedure in GR, by the newtonian limit
for $ r \rightarrow \infty, $  in terms of the Newton constant and
the mass of the compact object that is, $r_{H} = 2 g_{N} M/c^{2}.$
Thus, the final form of the effective metric is given by
\begin{eqnarray}
ds^{2} &=& ( 1 - \frac{r_{H}}{r} )  dT^{2} \nonumber \\
&-& \left( 1 - \frac{r_{H}}{r} \right)^{-1} \, \left[ 1 + \sigma^{2}
\,  \left(\frac{r_{H}}{r}
 \right)^{2}  \right] \,  dr^{2} \nonumber \\
 &-& r^{2} d\theta^{2} - r^{2} sin^{2}\theta d\varphi^{2}, \label{1setembro1435}
\end{eqnarray}
where $$ \sigma^{2} \equiv \frac{( \beta - 1)^{2}}{4 \beta}.$$

 It is a remarkable consequence of the above solution
 that in the case in which the self-interaction of the fundamental
 spinors vanishes and only the interaction between $\Psi$ and $\Upsilon$
 occurs, that is, for $\beta = 1$ the
 four-geometry is precisely the same as the
Schwarzschild solution in GR. On the other hand, if $\beta \neq 1$
the difference between both theories appears already in the order
$ ( r_{H}/r)^{2}.$ Indeed for General Relativity we have
\begin{equation}
 - g_{11} = 1 + \frac{r_{H}}{r} + ( \frac{r_{H}}{r} )^{2}
\nonumber
\end{equation}
and for the Spinor Theory we obtain \begin{equation*}
 - g_{11} = 1 + \frac{r_{H}}{r} + ( \frac{r_{H}}{r})^{2} \, ( 1 + \sigma^{2} ).
 \end{equation*}
The parameter $\beta$ should be fixed by observation.

\section{Conclusion}
In the present paper we have presented a new formalism to describe
gravity. We have shown that there is an alternative way to
implement the Equivalence Principle in which the geometry acting
on matter is not an independent field, and as such does not posses
its own dynamics. Instead, it inherits one from the dynamics of
two fundamental spinor fields $\Psi$ and $\Upsilon$ which are
responsible for the gravitational interaction and through which
the effective geometry appears. We have presented a specific model
by using Heisenberg equation of motion for the self-interacting
spinors. This equation of motion can be understood in terms of a
modification of the internal connection as seen by $\Psi$ and
$\Upsilon$ and only by these two spinors. This dynamics, which
involves not only the self terms but also a specific coupling
among these two fields, provides an evolution for the effective
metric (constructed in terms of these spinors) which is the way
the fields $\Psi$ and $\Upsilon$ interact with all other forms of
matter and energy. We have succeeded in finding a solution for the
fields, and thus we can extract the behavior of the effective
metric in the case of a static spherically symmetric
configuration. The result is similar as in General Relativity,
showing the existence of horizon and the possibility of existence
of Black Hole. They are not identical and differ already on the
order $ (r_{H}/r )^{2}.$

\section*{Acknowledgements}
I would like to thank my colleagues of ICRA-Brasil and particularly
Dr J.M. Salim with whom I exchanged many discussions concerning the
spinor theory of gravity and Dr L A R Oliveira for his comments and
suggestions  on the present paper. I would like to thank Dr Samuel
Senti for his kind help in the final english version of this
manuscript. This work was partially supported by {\em Conselho
Nacional de Desenvolvimento Cient\'{\i}fico e Tecnol\'ogico} (CNPq)
and {\em Funda\c{c}\~ao de Amparo \`a Pesquisa do Estado de Rio de
Janeiro} (FAPERJ) of Brazil.


\begin{thebibliography}{100}
\bibitem{GPPdesitter} In \cite{GPP} another choice is investigated, concerning the
deSitter geometry.
\end{thebibliography}
\end{document}